\documentclass[12pt,final]{iopart}
\pdfoutput=1

\usepackage{graphicx}
\usepackage[center]{subfigure}

\newcommand{\deriv}[2]{\ensuremath{\frac{\partial #1}{\partial #2}}}

\newcommand{\Vec}[1]{\ensuremath{\mathbf{#1}}}
\newcommand{\bvec}{\Vec{b}}
\newcommand{\kvec}{\Vec{\kappa}}
\newcommand{\bxk}{\bvec_0\times\kvec_0\cdot\nabla}

\newcommand{\Jpar}{J_{||}}

\newcommand{\delp}{\nabla_\perp^2}

\begin{document}

\title[Simulation of ELMs using BOUT++]{Simulation of Edge Localised Modes using BOUT++}
\author{B.D.Dudson$^1$, X.Q.Xu$^2$, M.V.Umansky$^2$, H.R.Wilson$^1$, P.B.Snyder$^3$}
\address{$^1$ Department of Physics, University of York, Heslington, York YO10 5DD \ead{bd512@york.ac.uk}}
\address{$^2$ Lawrence Livermore National Laboratory, Livermore, CA 94551, USA} 
\address{$^3$ General Atomics, San Diego, CA 92186, USA}

\begin{abstract}
The BOUT++ code is used to simulate ELMs in a shifted circle equilibrium.
Reduced ideal MHD simulations are first benchmarked against the linear ideal MHD code
ELITE, showing good agreement. Diamagnetic drift effects are included finding the expected
suppression of high toroidal mode number modes. Nonlinear simulations are performed, making the assumption
that the anomalous kinematic electron viscosity is comparable to the anomalous electron
thermal diffusivity. This allows simulations with realistically high
Lundquist numbers ($S = 10^8$), finding ELM sizes of 5-10\% of the pedestal stored thermal energy. 
Scans show a strong dependence of the ELM size resistivity at low Lundquist numbers,
with higher resistivity leading to more violent eruptions. At high Lundquist numbers relevant
to high-performance discharges, ELM size is independent of resistivity as hyper-resistivity
becomes the dominant dissipative effect.
\end{abstract}


\section{Introduction}

Future tokamak devices such as ITER and DEMO require high performance plasma operation
in order to demonstrate economical fusion performance. For this reason, discharges with
a transport barrier close to the plasma edge (H-mode \cite{keilhacker1984}) are
the current baseline operating scenario for ITER \cite{doyle2008}. 
Whilst transport barriers at the plasma edge result in improved performance, the steep pressure gradients
and associated bootstrap current can destabilise peeling-ballooning
modes \cite{snyder-2002}. These ideal MHD instabilities are thought
to be responsible for triggering the observed eruptions of filamentary structures from the plasma edge 
known as Edge Localised Modes (ELMs). The particles and energy released during ELMs are
deposited on material surfaces and are potentially damaging in future devices.
There is therefore interest in understanding and controlling these events.

Understanding of the linear onset and structure of peeling-ballooning modes is now
quite well developed, with codes such as ELITE~\cite{snyder-2002,wilson-2002}, GATO~\cite{bernard1981} and
MISHKA~\cite{mikhailovskii1997,chapman2006}
able to predict experimental operating limits\cite{snyder-2002,hegna-1996,connor-1998}. 
Study of the nonlinear evolution of these instabilities is more challenging, and
although there are analytic theories \cite{wilson-2004}
and semi-analytic models \cite{zhu-2006,zhu-2007} for the early non-linear evolution,
it is not yet fully understood how this develops into the filamentary structures observed,
and ultimately how particles and energy are lost.

Several 3D non-linear codes have been used to simulate ELMs, including NIMROD 
\cite{sovinec-2004,brennan-2006,pankin-2007,burke2010},
BOUT \cite{snyder-2005,snyder-2006}, JOREK \cite{huysmans-2007,huysmans2009,pamela2010}, 
GEM \cite{scott-2005,scott-2006}, M3D \cite{park-1999,sugiyama-2006,sugiyama2009} and M3D-C1~\cite{Ferraro2009}.
These codes incorporate a wide range of physics
including (in the case of BOUT and some NIMROD simulations) 2-fluid effects. 
In this paper, simulations of ELMs using the the BOUT++ code~\cite{Dudson2009} based on modified
reduced MHD in a shifted circle equilibrium are presented, expanding on and extending results 
presented in \cite{xu2010}. By introducing a hyper-resistive
term into Ohm's law, realistic resistivities have been simulated, and the effect of varying
resistivity on ELM sizes has been investigated. 

Section~\ref{sec:model} describes the starting set of equations used for comparison with 
linear ideal MHD, and the equilibrium simulated is described in section~\ref{sec:equil}. A novel
means to handle the vacuum region which has been found to work well in ideal or resistive MHD
simulations is described in section~\ref{sec:vac}. Linear simulations are benchmarked against
linear ideal MHD codes in section~\ref{sec:elm_lin}, and nonlinear simulations of ideal reduced MHD
and the issues encountered are discussed in section~\ref{sec:nlideal}. By incorporating diamagnetic
drift and either high resistivity or a hyper-resistivity, simulations of nonlinear eruptions are performed
and discussed in sections~\ref{sec:diaresist} and \ref{sec:diahyper} respectively. The effect of
varying resistivity on ELM crash size is then discussed in more detail in section~\ref{sec:elmsize}.
Details of the magnetic field structure are presented in section~\ref{sec:magfield}, indicating a
reconnection
of flux-surfaces at the leading edges of the erupting filaments. 

\section{Model equations}
\label{sec:model}

In this paper, the starting equations are those of high-$\beta$ reduced MHD \cite{hazeltine-2003},
evolving pressure $P$, vorticity $U$ and magnetic potential $\psi = A_{||}/B_0$:
\begin{eqnarray}
\rho \frac{dU}{dt} &=& B^2\bvec\cdot\nabla\left(\frac{\Jpar}{B}\right) + 2\bxk P \label{eq:vort} \\
\deriv{\psi}{t} &=& -\frac{1}{B_0}\nabla_{||}\phi \label{eq:psi} \\
\deriv{P}{t} &=& -\frac{1}{B_0}\bvec_0\times\nabla\phi\cdot\nabla P \label{eq:pressure} \\
U &=& \frac{1}{B_0}\delp\phi \label{eq:vorticity_def} \\
\Jpar &=& J_{||0} - \frac{1}{\mu_0}B_0\delp\psi \label{eq:jpardef}
\end{eqnarray}
where $\nabla_{||}F = B\partial_{||}\left(F/B\right)$ for any $F$,
$\partial_{||} = \left(\mathbf{b}_0 + \mathbf{b}_1\right)\cdot\nabla$, $\mathbf{b}=\mathbf{B}/B$.
The plasma density is constant in space and time, fixed at $10^{19}$m$^{-3}$.
The perturbed magnetic field is given by $\mathbf{B}_1 =  \nabla\psi\times\mathbf{B}_0$ where $\mathbf{B}_0$
is the unperturbed field.
$\kappa_0 = \mathbf{b}_0\cdot\nabla\mathbf{b}_0$ is the field curvature
and all quantities are in SI units. The vorticity equation~\ref{eq:vort} includes
the kink/peeling term through the perturbed magnetic field: 
\[
\bvec\cdot\nabla\left(\frac{\Jpar}{B_0}\right) = \left(\bvec_0\cdot\nabla - \bvec_0\cdot\nabla\psi\times\nabla\right) \left(\frac{\Jpar}{B_0}\right)
\]
This minimal set of equations contains the basic physics needed to describe peeling-ballooning
modes, including pressure/curvature and parallel current (kink/peeling) instability drives, and
field-line bending stabilisation. The model is a useful starting point because
it allows benchmarking against linear ideal MHD codes such as ELITE, providing a valuable means
of checking the results of these simulations. 

In the pedestal with steep pressure gradients diamagnetic
effects are expected to be significant, and damp high toroidal mode-number,$n$, instabilities which can
lead to problems in ideal MHD simulations (see section~\ref{sec:nlideal}). Therefore, the set of equations
(\ref{eq:vort}-\ref{eq:jpardef})
is modified in section~\ref{sec:diaresist} to include the diamagnetic drift.

Dirichlet (zero value) boundary conditions are used for the perturbed pressure, vorticity and
parallel current. To be consistent, the boundary conditions for $\psi$ and $\phi$ must satisfy
$\nabla_\perp^2\psi = 0$ and $\nabla_\perp^2\phi = 0$. This is done by solving for each toroidal
mode-number which have solutions $\hat{\phi}_n \simeq C_+\exp\left(\sqrt{g^{\zeta\zeta}/g^{\psi\psi}}k_\zeta x\right) + C_-\exp\left(-\sqrt{g^{\zeta\zeta}/g^{\psi\psi}}k_\zeta x\right)$ where $\zeta$ is the toroidal angle,
$k_\zeta = n/R$ and $x$ is the radial ($\psi$) coordinate.
Only solutions which decay going out of the domain are allowed, so $C_+=0$ on the outer boundary,
and $C_-=0$ on the inner boundary.

\subsection{Vacuum region}
\label{sec:vac}

A complication of simulating the plasma edge region is the handling of the
vacuum region, and the plasma-vacuum interface. In linear codes such as ELITE an analytical
approach can be used; non-linear codes must also handle motion of the
vacuum interface. The usual approach has been
to treat the vacuum as a resistive plasma, with a jump in resistivity between
plasma and vacuum~\cite{burke2010}. Here a different approach is used: in the vacuum we evolve
$\psi$ towards a self-consistent solution determined by currents in the core only. The procedure
is as follows.

First define a smoothed step function $\Theta$, which switches between $0$
in the core to $1$ in the vacuum:
\begin{equation}
\Theta = \frac{1}{2}\left[1-\tanh\left(\frac{P - P_{vac}}{\Delta P_{vac}}\right)\right]
\label{eq:vacmask}
\end{equation}
where $P_{vac}$ is the pressure at the plasma-vacuum interface, and $\Delta P_{vac} < P_{vac}$
is the transition width. 
At any given time, the solution
$\psi\left(\mathbf{x},t\right)$ gives a current $\Jpar^{sol}$
\[
\Jpar^{sol} = -\frac{1}{\mu_0}B_0\delp\psi
\]
which may or may not include currents in the vacuum region. From this, a ``target''
current $\Jpar^{target}$ is calculated by setting all vacuum currents to zero
\[
\Jpar^{target} = \Jpar^{sol} \left(1 - \Theta\right)
\]
This is the closest physically acceptable solution to the current result. From this, 
a target $\psi$ can be calculated
\[
\psi^{target} = \nabla_\perp^{-2}\left(\mu_0 \Jpar^{target} / B_0\right)
\]
In the vacuum region, this is the solution which would give zero current, but simply
setting $\psi$ to this value in the vacuum would result in an inconsistency because
$\psi$ in the core is affected by currents in the vacuum. Instead,
equation~\ref{eq:psi} is modified to
\[
\deriv{\psi}{t} = -\left(1-\Theta\right)\frac{1}{B_0}\nabla_{||}\phi +  \Theta\left(\psi^{target} - \psi\right) / \tau_{jvac}
\]
and so $\psi$ converges on the target value with a small time-constant
$\tau_{jvac}$. $\psi$ therefore evolves to a self-consistent state with
zero current in the vacuum region.

This method has been found to work well for ideal and resistive simulations, even in the nonlinear
regime. It is useful because this method uses the same code already used for solving the reduced MHD model
with minimal modifications. At this point this method does not work once diamagnetic effects are included
and so is not used in the nonlinear results presented in sections~\ref{sec:diaresist} and
\ref{sec:diahyper}.

\section{Equilibrium}
\label{sec:equil}

Simulation of ELMs in full x-point geometry is necessary in order to predict the behaviour of ELMs
in future devices. As a starting point however, it is useful to use equilibria which remove many 
of the complications of real magnetic geometry: This simplifies study of the basic physics of ELMs, but more
importantly enables accurate benchmarking of the results against linear theory and other codes.

\begin{figure}[htbp!]
\centering
\subfigure[Pressure and $J_{||}$ profiles]{
  \label{fig:profiles}
  \includegraphics[scale=0.7]{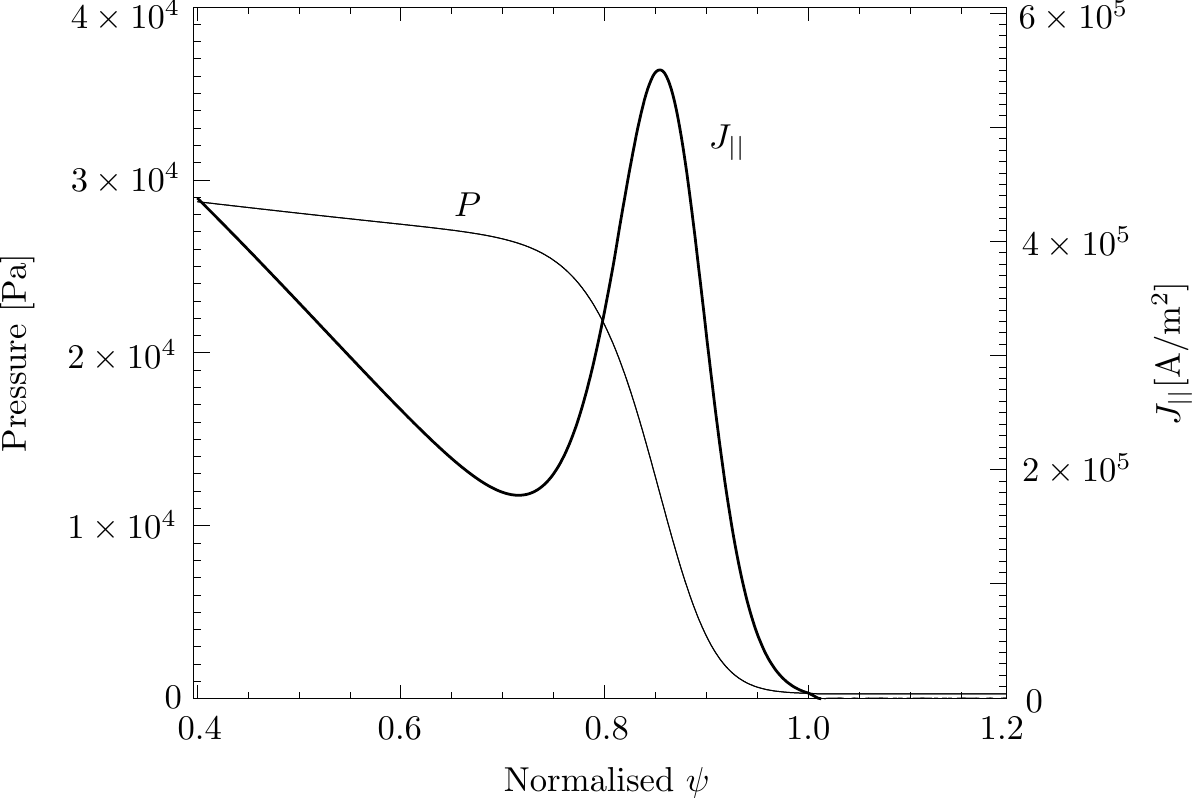}
}
\subfigure[Poloidal cross-section. Radial mesh coarsened for clarity]{
  \label{fig:mesh}
  \includegraphics[scale=0.75]{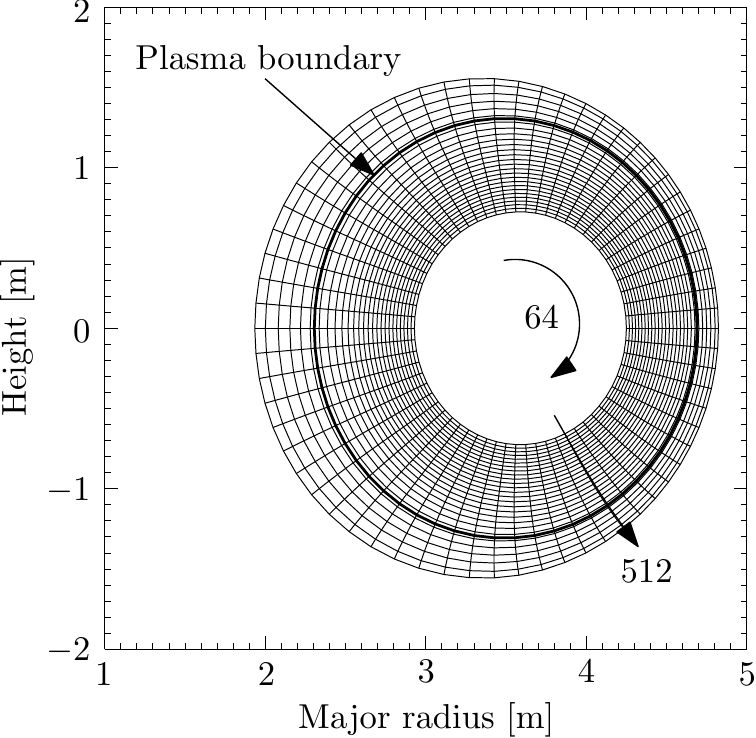}
}
\caption{Equilibrium (cbm18\_dens8) profiles and mesh}
\label{fig:cbm18}
\end{figure}

A test case which is becoming standard in this field is the \texttt{cbm18} series, created 
by P.Snyder using the TOQ equilibrium code. The case used here \texttt{cbm18\_dens8} has now been
used by NIMROD~\cite{burke2010} and M3D-c1~\cite{Ferraro2009} to benchmark linear growth-rates against the
ELITE~\cite{snyder-2002,wilson-2002} and GATO~\cite{bernard1981} linear MHD stability codes.
The pressure and current profiles are shown in figure~\ref{fig:profiles} over the range 
of normalised $\psi$ simulated using BOUT++ ($0.4<\psi_n<1.2$). 
The definition of the plasma
edge is somewhat arbitrary since here there are no separatrices, and so $\psi_n=1$ is defined as
the point where the equilibrium plasma pressure gradient falls to zero. For this equilibrium, the normalised beta
is $\beta_N=1.51$,
edge $q_a=3.03$, pedestal toroidal pressure $\beta_{t0}=1.9\times 10^{-2}$ and normalised pedestal
width is $L_{ped}/a=0.049$.

The mesh used for the simulations in this paper is shown in figure~\ref{fig:mesh}. This has
a resolution of 512 points in radial coordinate $\psi$ and 64 in poloidal coordinate $\theta$.
In the toroidal direction
64 points are usually used, with the results checked by doubling this to 128. Note that despite 
this relatively low resolution in $\theta$, high $m$ numbers can be simulated. This is
because the field-aligned coordinate system
\[
x = \psi - \psi_0 \qquad y = \theta \qquad z = \phi - \int_{\theta_0}^\theta\nu\left(\psi,\theta\right)d\theta
\]
(where $\nu\left(\psi,\theta\right)$ is the field-line pitch) means that the $\theta$ direction is aligned
with the magnetic field and so $n_\theta = 64$ is therefore the number of points along a field-line per
poloidal turn. The effective poloidal resolution depends on the toroidal resolution (number of field-lines simulated $\times$ number of repetitions in a full torus) and the local safety factor $q\sim q_a=3.03$, and so is 
$\simeq 3\times 64 \times 5= 960$ for the simulations shown here.

In order to avoid problems associated with magnetic shear and deformation of grid cells associated
with field-aligned coordinate systems, shifted local coordinate systems~\cite{dimits-1993,scott-2002}
are used so that the radial derivative is always taken in $\psi$. Further details of this coordinate
system and its implementation in BOUT++ can be found in \cite{Dudson2009}.

\section{Linear benchmarking}
\label{sec:elm_lin}

\begin{figure}[htb!]
\centering
\includegraphics[scale=0.7]{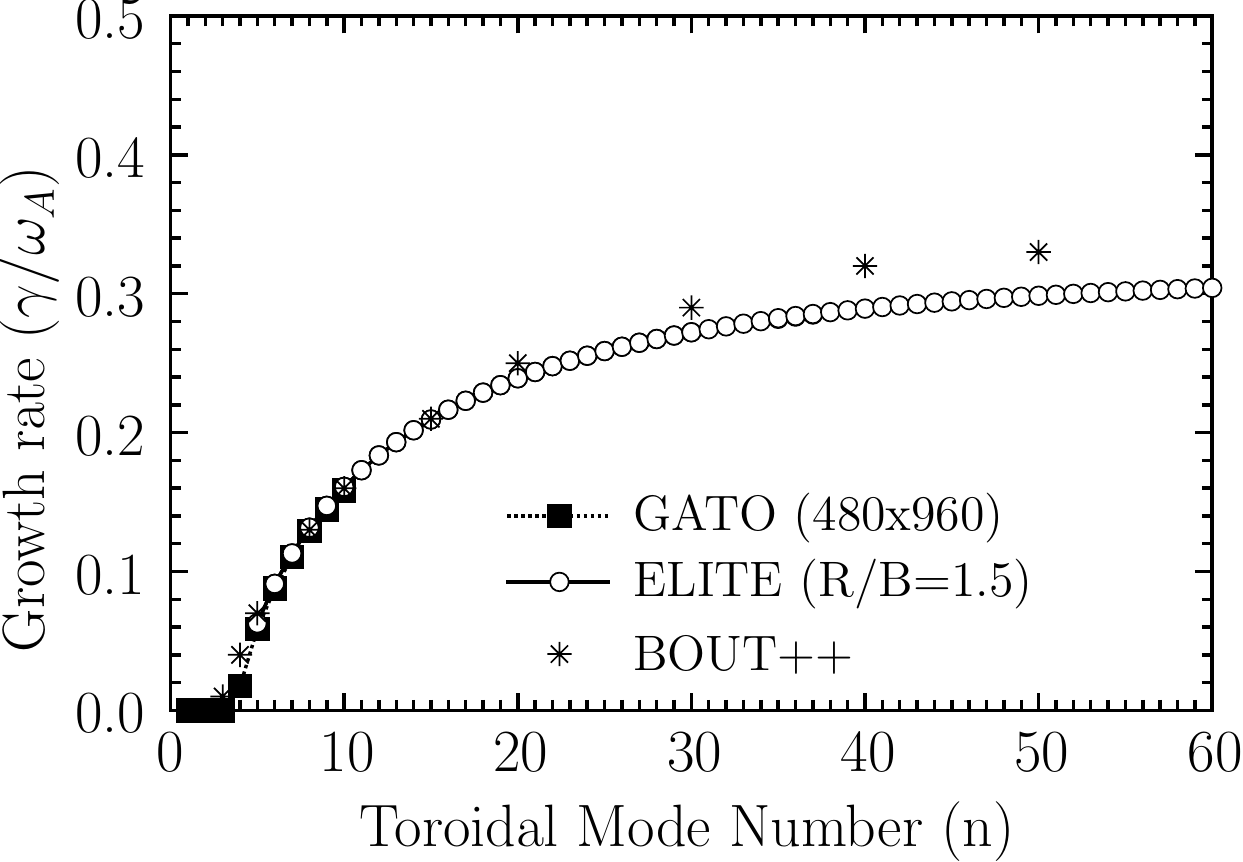}
\caption{Linear growth rate of a strongly ballooning test case, as calculated by BOUT++
and the linear ideal MHD codes ELITE and GATO}
\label{fig:cbm18_8_gr}
\end{figure}

In order to benchmark BOUT++, linear simulations have been
performed and comparison made to the ELITE~\cite{snyder-2002,wilson-2002} and
GATO~\cite{bernard1981} linear ideal MHD codes.
Figure~\ref{fig:cbm18_8_gr} shows the linear growth-rate from BOUT++ for this equilibrium (star symbols), along
with the results from ELITE (open circles) and GATO (filled squares).
All growth-rates are normalised to an Alfv\'en frequency
$\omega_A = V_A / R$ , with $V_A$ the  Alfv\'en velocity, and $R$ the major radius.
For all the simulations presented in this paper, a reference density of $10^{19}$m$^{-3}$
was used, so that time is normalised to $\tau_A = 0.37\mu$s.

At high-$n$, the BOUT++ result begins to deviate from the ELITE result, whereas it might
be expected that finite-$n$ corrections would lead to the greatest differences being at low $n$.
The reason for this is that as the mode number $n$ is increased the radial width of the mode becomes
narrower: Since these simulations all used the same radial resolution, at high $n$ small-scale
structures can no longer be resolved properly. The mode structure of the radial displacement $\xi_\psi$
from ELITE and BOUT++ have also been compared \cite{Dudson2009}, finding good agreement.
Individual $\left(m,n\right)$ modes are found to peak close to their resonant magnetic surfaces, as is expected
from analytic theory.
This test is a proof-of-principle which demonstrates that BOUT++ is capable of simulating
the ideal ballooning mode correctly using ideal reduced MHD. Linear benchmarks are 
a necessary but not sufficient condition for non-linear results: comparisons against other nonlinear
codes and ultimately experimental results are needed to verify the fidelity
of nonlinear results.

\section{Non-linear ideal MHD}
\label{sec:nlideal}

In ideal MHD linear simulations, peaks in the parallel current form at rational $q$ surfaces. The width
of these current sheets is of the order of the distance between rational surfaces
$\Delta_q \sim r/nqs \simeq 6$mm, which can be resolved in the linear regime since 
when using 512 radial points across the pedestal the grid spacing is $\sim 1.1$mm.
Linear codes such as ELITE and MARS~\cite{Bondeson1992}
typically pack mesh-points around rational surfaces to handle these features, but this becomes
much more challenging in the nonlinear regime: flux-surfaces move over time, current sheets
can be compressed so reducing scale lengths, and any adaptive scheme will
struggle to handle highly nonlinear regimes where flux surfaces may be destroyed.

Nonlinear simulations of ideal MHD have been attempted using BOUT++, but as yet without success: As
flux-surfaces are distorted, small-scale structures are generated in the binormal direction which corresponds to the generation of high toroidal mode-numbers.
The size of the structures formed in the toroidal direction is approximately given by
rotating the radial scales in the perpendicular plane, and projecting in the toroidal direction
which gives $L_\zeta\simeq \Delta B_\zeta / B_\theta \sim 5$cm and so a maximum toroidal mode-number of
$n_{max}\sim 600$.

A second source of high-$k$ structures and hence numerical problems is that the
modes with the highest growth-rate are those with the highest $n$ (i.e. those
at the grid scale). Any perturbation at high $n$ will therefore rapidly grow and eventually
dominate the solution. 

Handling small-scale structures is a problem in many nonlinear simulations, and if
possible then these should be resolved. In practice however, it is often necessary to
add some sort of numerical dissipation to remove structures at the grid scales.
Several approaches are commonly used to achieve this. For example the solution can be
filtered in Fourier space (so only the low-$n$ modes are simulated) or a hyper-diffusion
like term can be added. Whichever method is used, it should be chosen to minimise the impact
on the large-scale structures of interest. 

Tokamak simulations have the advantage that Larmor radius effects should naturally 
damp high-$k$ structures. Adding $\rho_i$ effects such as diamagnetic drifts both
make the simulation more realistic and also aid numerical stability by damping high-$k$
structures. In this paper, nonlinear simulations are performed by first adding diamagnetic drift
and following the common practice of incorporating a high resistivity in section~\ref{sec:diaresist},
before introducing a hyper-resistive term in section~\ref{sec:diahyper}.

\section{Diamagnetic and resistive simulations}
\label{sec:diaresist}

For these initial nonlinear simulations, the basic effects of diamagnetic stabilisation
are incorporated into the model by modifying the vorticity (\ref{eq:vorticity_def})
so that the total plasma velocity is now given by the sum of $E\times B$ and diamagnetic
drifts.
\[
\mathbf{V}\simeq \mathbf{V}_E + \mathbf{V}_D = \frac{1}{B^2}\mathbf{B}\times\nabla\phi + \frac{1}{enB^2}\mathbf{B}\times\nabla P
\]
Hence assuming constant mass density $n$, this gives
\begin{equation}
U = \frac{1}{B}\delp\left(\phi + \frac{P}{en}\right)
\label{eq:vorticity_dia}
\end{equation}
In this simplified model, both equilibrium and turbulent zonal ($m=n=0$) flows have been set
to zero: $\mathbf{V}_{E0} + \mathbf{V}_{D0} = 0$ and toroidally-averaged $\left<\delta\mathbf{v}_E +
\delta\mathbf{v}_D\right> = 0$. The equilibrium radial electric field is therefore given by
$E_{r0}=\left(1/n_0Z_ie\right)\nabla_r P_{i0}$ (with $P_{i0}=P_0/2$) and perturbed radial electric
field $\left<E_r\right>=\left(1/n_0Z_ie\right)\nabla_r \left<P_i\right>$. Recent work 
\cite{huysmans2009,pamela2010,strauss1995} indicates that the nonlinear formation of zonal flows results in multiple
filaments breaking off from the plasma, but investigation of this effect is beyond the scope of this
paper and the subject of future work.

At realistic resistivities of $S\sim 10^8$, the scale of the current sheets is limited by resistivity to
approximately $\Delta_R\simeq R\sqrt{\left(\omega_A/\gamma\right)/S}\sim 10-100\mu$m, of the same order as the 
electron gyro-radius. The radial grid spacing in these simulations $\Delta_\psi \simeq 1$mm, and so
to limit the smallest scales to these sizes a greatly increased resistivity of $S=10^4$ is used,
constant in space.

\begin{figure}[htb!]
\centering
\includegraphics[]{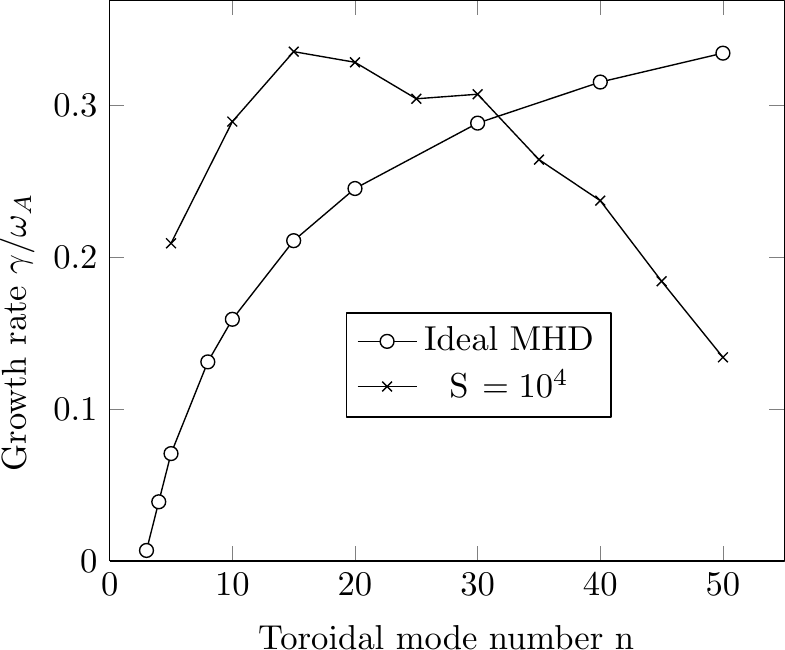}
\caption{Linear growth rate for simulations with $\omega_*$ and $S=10^4$}
\label{fig:resistgrow}
\end{figure}

\begin{figure}[htbp!]
\centering
\subfigure[Plasma displacement at the outboard mid-plane outwards (solid) and inwards (dashed)]{
  \label{fig:disp01}
  \includegraphics[scale=0.4]{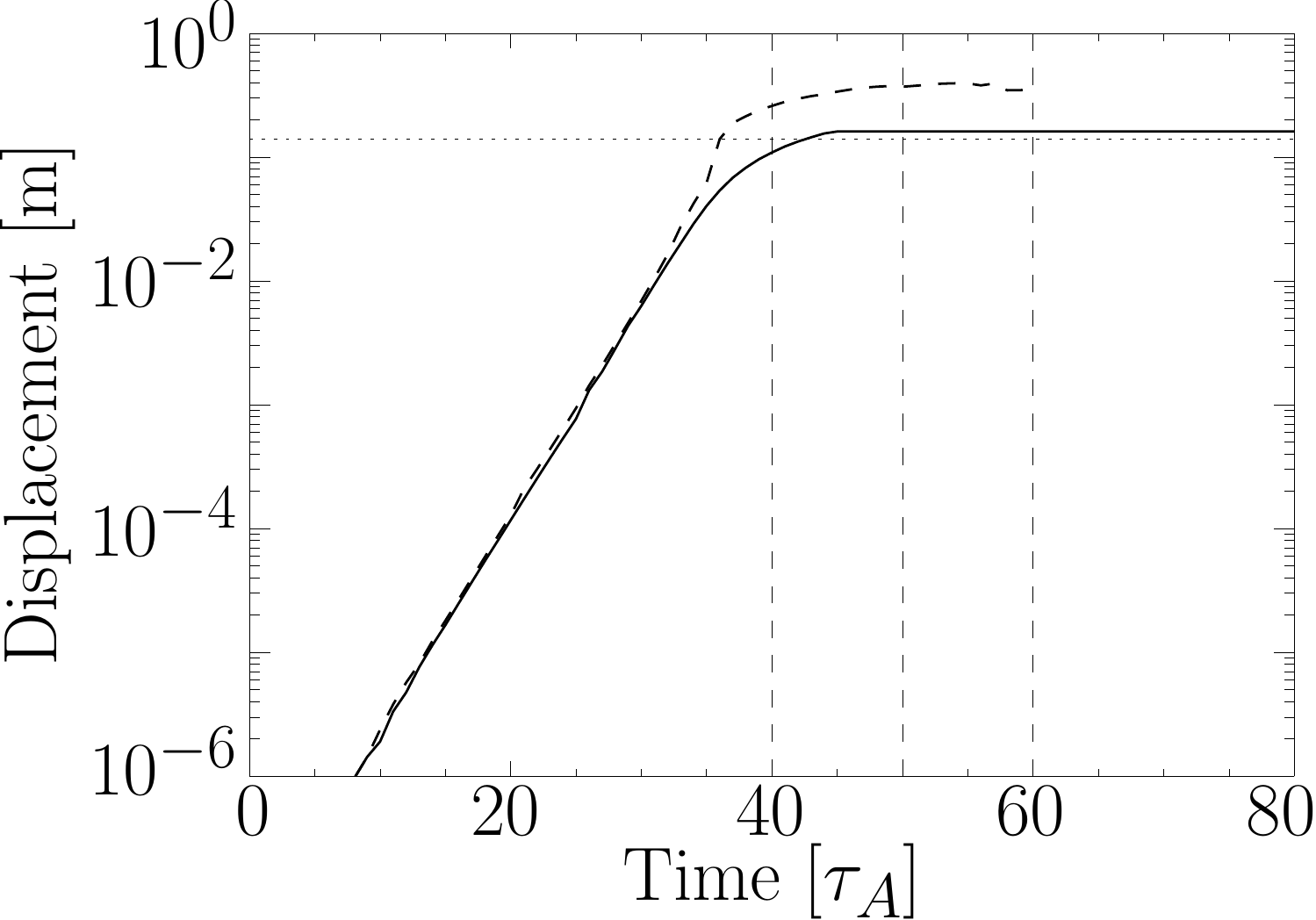}
}
\subfigure[Surface-averaged pressure profiles]{
  \label{fig:pavg01}
  \includegraphics[scale=0.4]{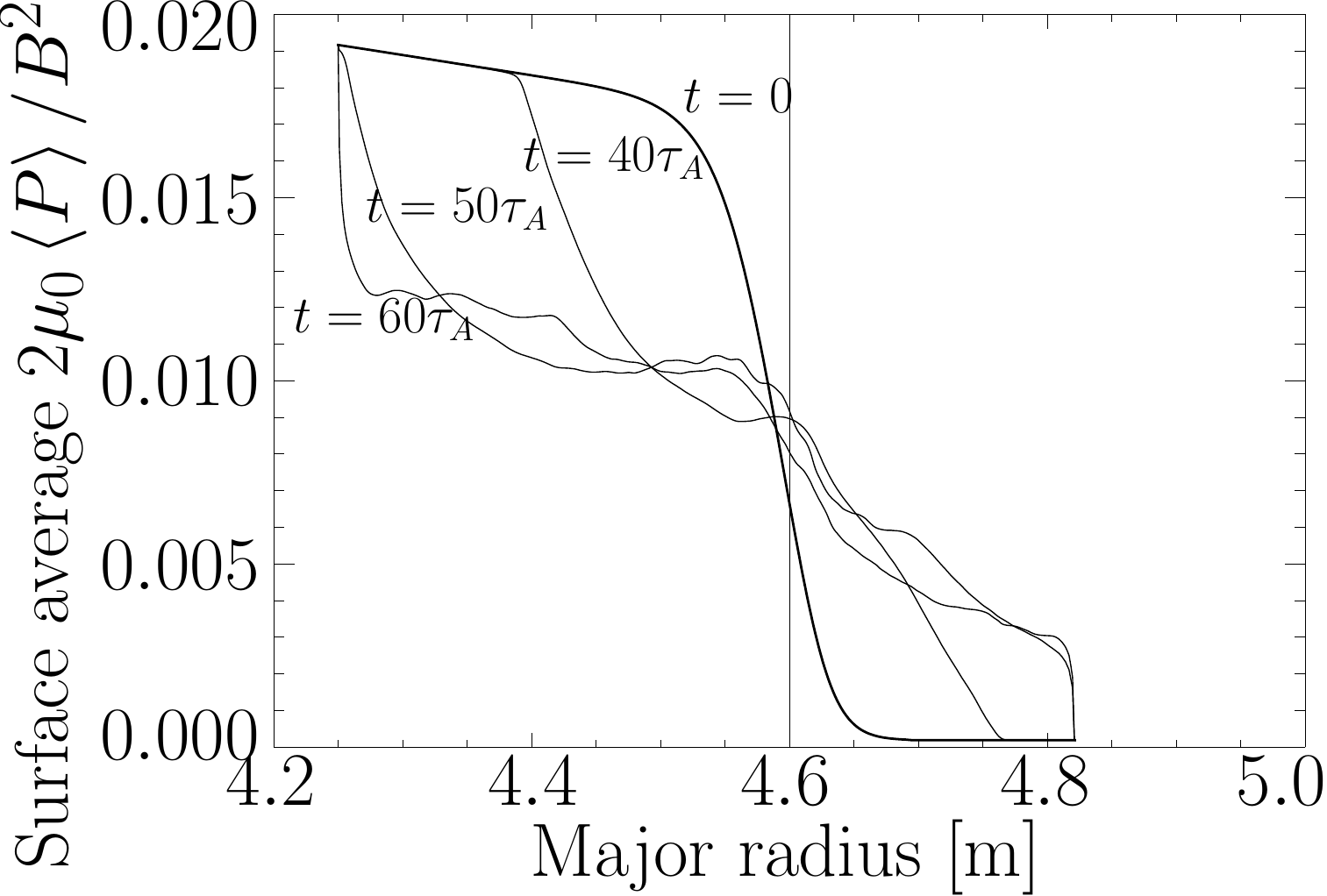}
}
\subfigure[Plasma energy loss $\Delta W_{ped}/W_{ped}$]{
  \label{fig:wped01}
  \includegraphics[scale=0.4]{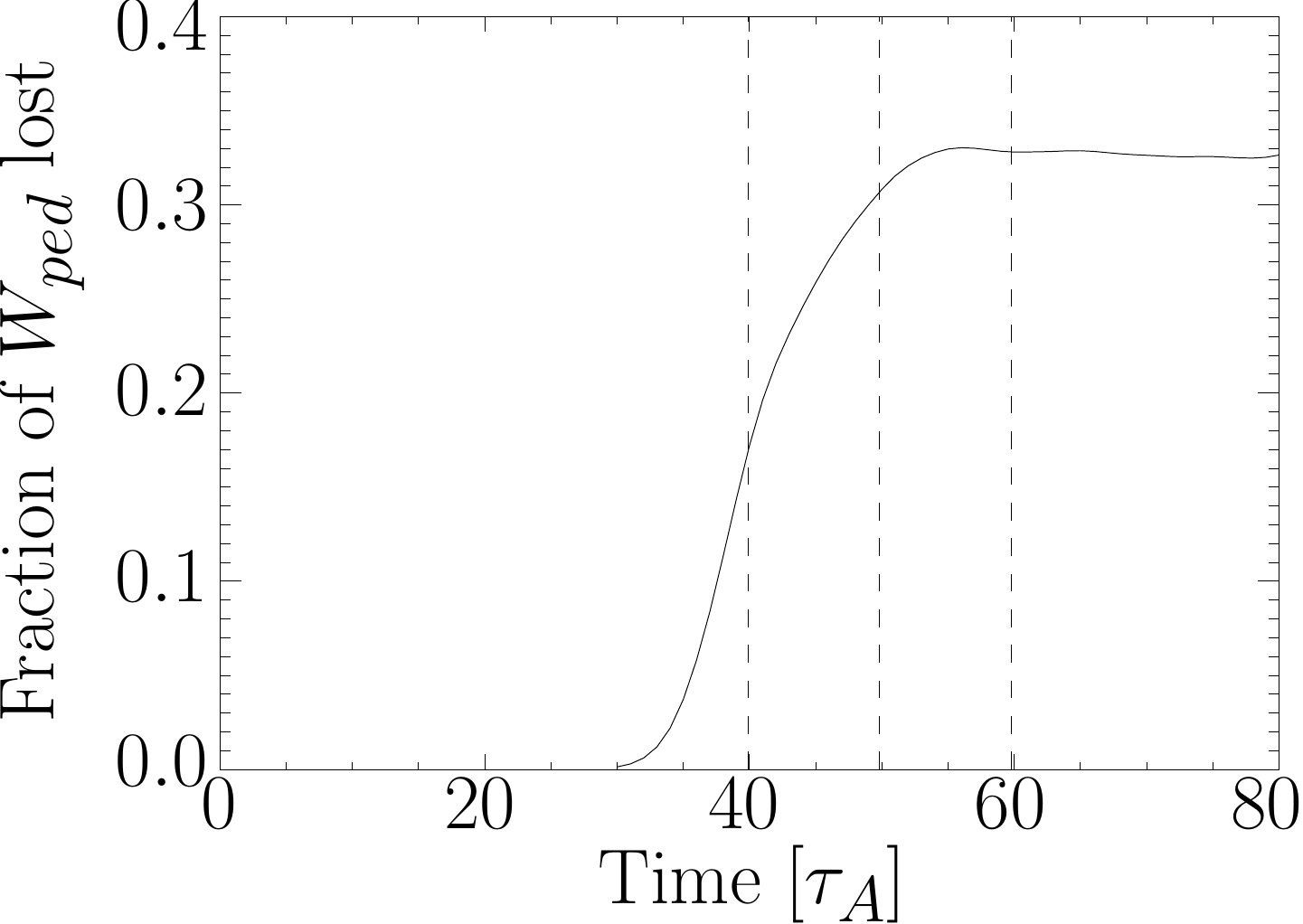}
}
\caption{Nonlinear simulation results with $\omega_*$ and $S=10^4$}
\label{fig:run_01}
\end{figure}

Diamagnetic drift and high resistivity produces the linear growth-rates shown in
figure~\ref{fig:resistgrow}.
High resistivity results in an increase in growth-rate above the ideal MHD case (open circles), whilst
high-$n$ modes are suppressed by diamagnetic drift effects in line with theoretical expectations. The
peak growth-rate now occurs at $n=15$, and so this toroidal mode-number is used as a starting point
for the nonlinear results shown in figure~\ref{fig:run_01}.
Figure~\ref{fig:disp01} shows the plasma displacement, measured by finding where the pressure
crosses a threshold. Outwards displacement is shown as a solid line, and inwards displacement as
a dashed line: These are approximately equal in the linear regime (as expected since mode goes like
$\exp\left(in\zeta\right)$ ), but begin to diverge once the displacement exceeds $\sim 1-2$cm with the
outgoing filament growth-rate reducing whilst the ingoing hole continues to grow at the linear rate. 

Figure~\ref{fig:pavg01} shows the pressure profiles averaged over (equilibrium) flux surfaces at
the times indicated by vertical lines in figure~\ref{fig:disp01}. This shows that this eruption propagates
far into the plasma beyond the steep pressure-gradient region
(shown in the $t=0$ equilibrium profile in figure~\ref{fig:pavg01}, and in figure~\ref{fig:profiles}).

To quantify the size of the ELM eruption, and provide a measure which can be compared against typical 
experimentally measured ELM sizes, the thermal energy from the inner edge of the domain ($R=4.2$m at the
outboard mid-plane) to $R=4.6$m (vertical line in figure~\ref{fig:pavg01}) $W_{ped} = \int 3P/2 dV$ is used.
The fraction of this energy lost during the ELM crash $\Delta W_{ped}/W_{ped}$ is shown in
figure~\ref{fig:wped01}. For this highly resistive case, approximately $30$\% of the thermal
energy in the pedestal is lost, limited by the radial size of the computational domain.


\section{Diamagnetic and hyper-resistive simulations}
\label{sec:diahyper}

Using high resistivities to damp small-scale structures and maintain numerical stability 
in the nonlinear regime results in greatly increased growth-rates, modified stability 
thresholds, and large eruptions which propagate far into the plasma. 

An alternative effect which damps high-$k$ structures whilst minimising the effect on large-scale
structures is hyper-resistivity \cite{xu2010,kaw1979,caunt-2001}: The parallel electric field
equation for $\psi = A_{||}/B_0$ is modified to
\[
\deriv{\psi}{t} = -\frac{1}{B_0}\nabla_{||}\phi + \frac{\eta}{\mu_0}\nabla_\perp^2\psi +\frac{\eta_H}{\mu_0}\nabla_\perp^4\psi
\]
where $\eta_H$ is a hyper-resistivity. The hyper-Lundquist number is then defined analogously to the Lundquist number as
the dimensionless ratio of an Alfv\'en wave crossing time to a hyper-diffusion timescale.
$S_H = \mu_0R_0^3v_A/\eta_H = S/\alpha_H$. The hyper-Lundquist parameter $\alpha_H = \eta_H/R^2\nu_{ei}$
can be estimated from electron collisional viscosity \cite{xu2010} as $\alpha_H \simeq \mu_e/R^2\nu_{ei}$ if it is
assumed that the electron viscosity $\mu_e$ is comparable to the anomalous electron thermal diffusivity
$\mu_e \simeq \chi_e \simeq 1$m$^2$/s. Taking a value of $\nu_{ei}\simeq 10^5$s$^{-1}$ gives $\alpha_H\simeq 10^{-4}-10^{-6}$ for $R\simeq 3$m.

Taking $S=10^8$ and $S_H=10^{12}$, the perpendicular scale-length of the parallel current can be
estimated this time by balancing $\partial\psi/\partial t$ against the hyper-resistive term. This gives
$\Delta_H\simeq R\left(\omega_A/\gamma/S_H\right)^{1/4}\simeq 2$mm, which is of the
order of the radial grid spacing $\Delta_\psi \simeq 1.1$mm, and an order of magnitude larger than was the
case with only resistivity included. By including this hyper-resistive effect, realistic resistivities
of $S\sim 10^8$ can be simulated.

\begin{figure}[htb!]
\centering
\includegraphics[]{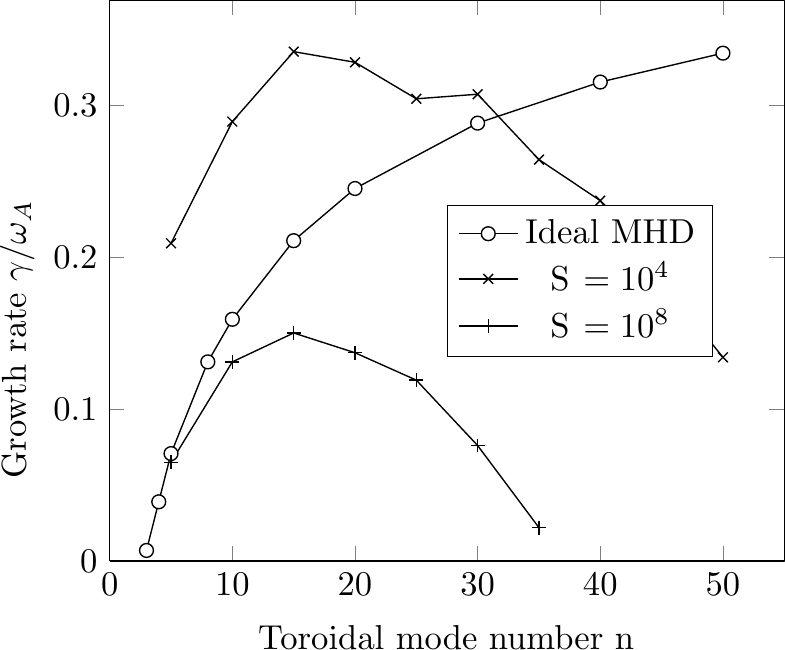}
\caption{Linear growth rate for simulations with $\omega_*$ and $S_H=10^{12}$}
\label{fig:hyperresistgrow}
\end{figure}
Linear growth-rates are shown in figure~\ref{fig:hyperresistgrow}, with the ideal MHD 
and resistive (figure~\ref{fig:resistgrow}) results for comparison. Growth-rates are 
below the ideal MHD case, stabilised at high-$n$ by diamagnetic drifts as expected, and recover
the low-$n$ ideal MHD growth-rates. This is encouraging and an improvement on the
resistive case where enhanced growth-rates were found.

\begin{figure}[htbp!]
\centering
\subfigure[Plasma displacement at the outboard mid-plane outwards (solid) and inwards (dashed)]{
  \label{fig:disp18}
  \includegraphics[scale=0.4]{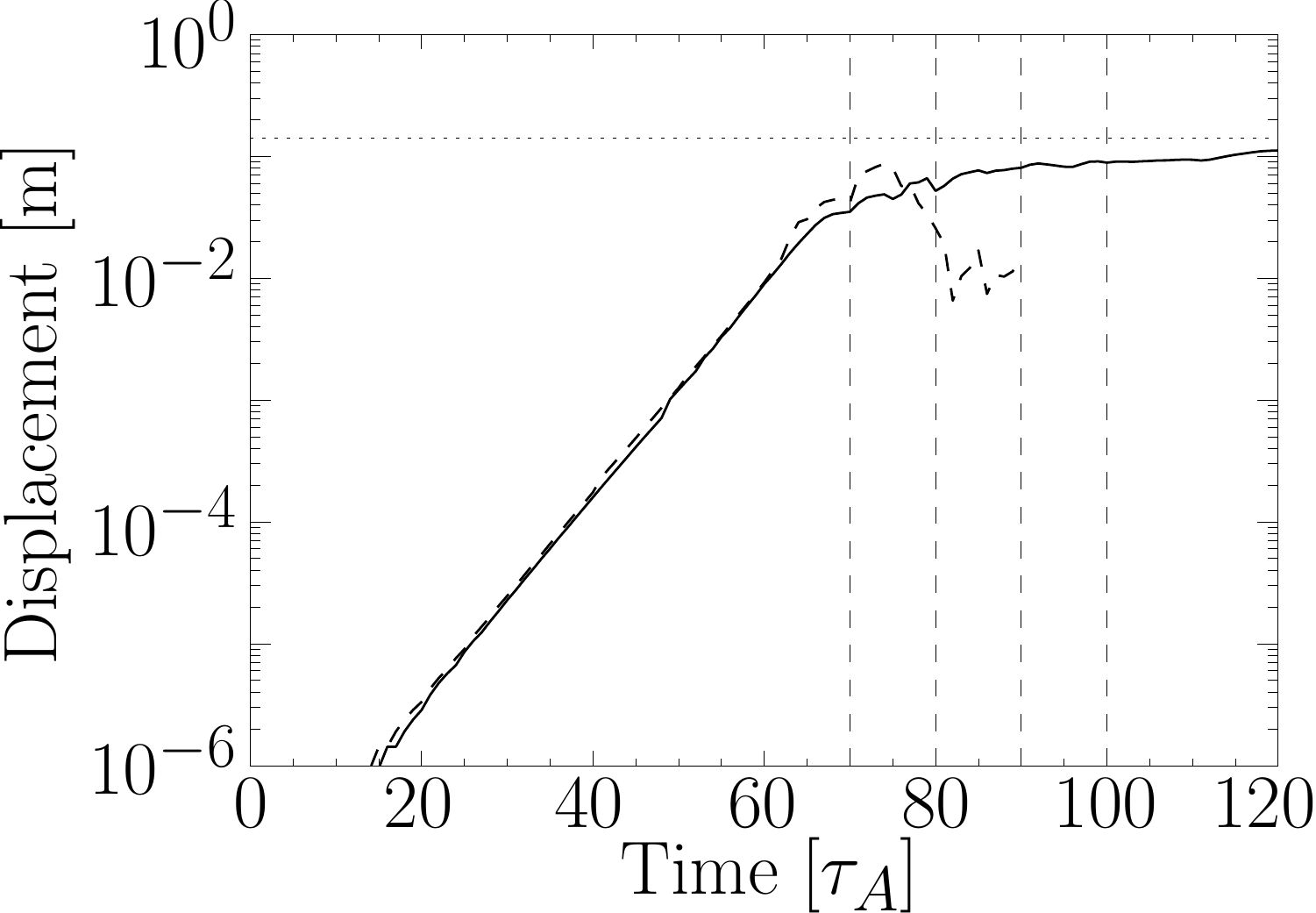}
}
\subfigure[Surface-averaged pressure profiles]{
  \label{fig:pavg18}
  \includegraphics[scale=0.4]{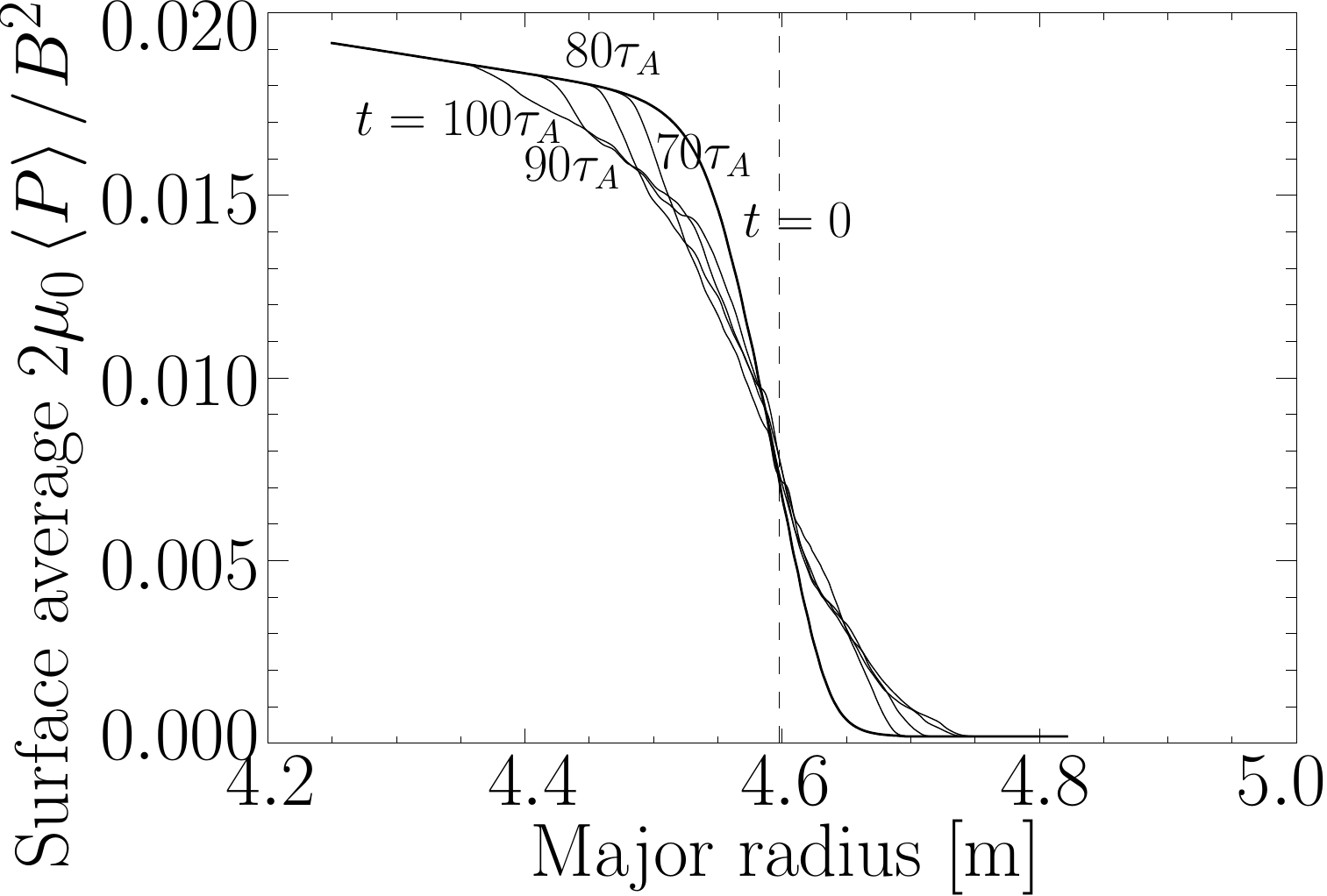}
}
\subfigure[Plasma energy loss $\Delta W_{ped}/W_{ped}$]{
  \label{fig:wped18}
  \includegraphics[scale=0.4]{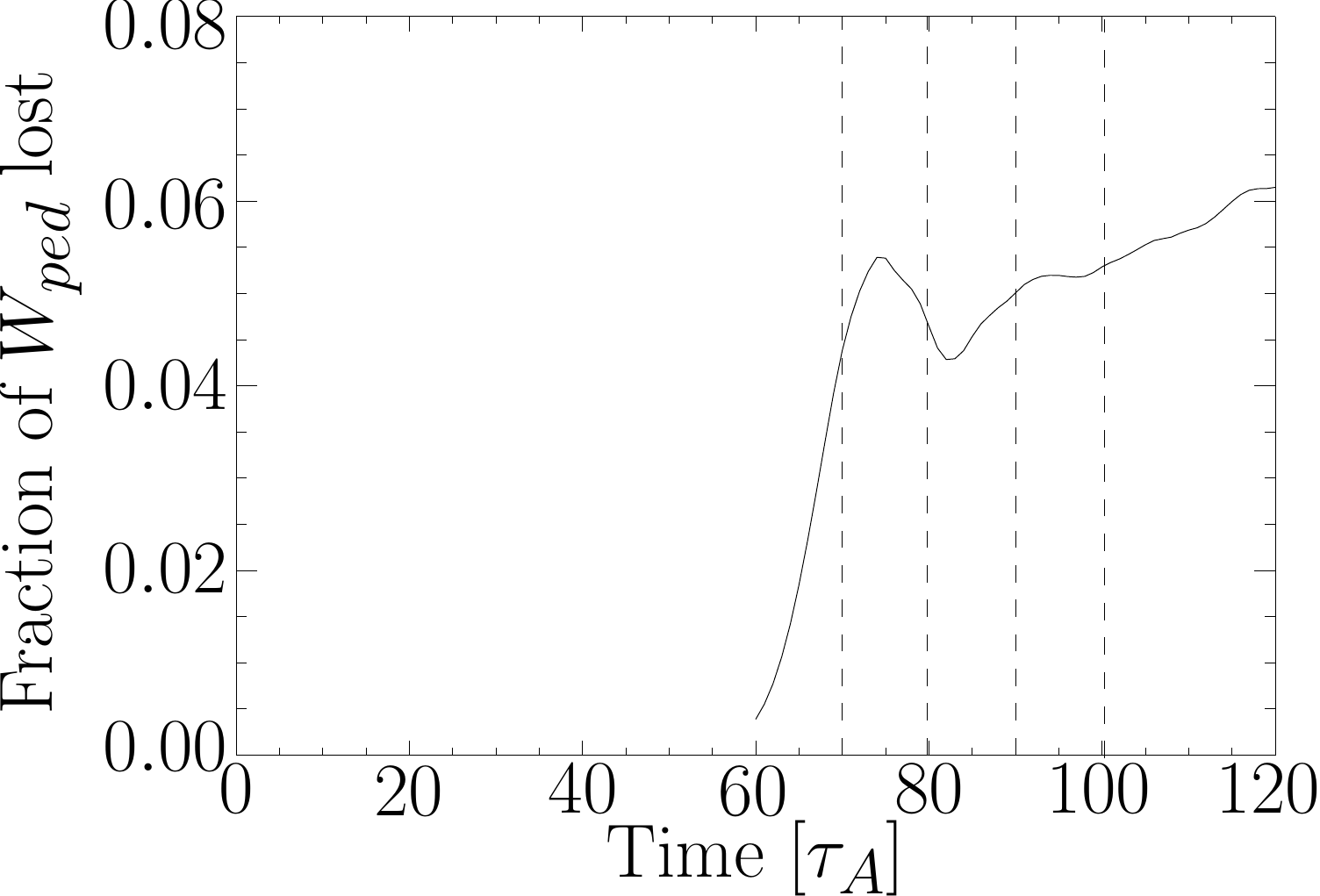}
}
\caption{Nonlinear simulation results with $\omega_*$, $S=10^8$ and $S_H=10^{12}$}
\label{fig:run_18}
\end{figure}

Nonlinear simulation results are shown in figure~\ref{fig:run_18} using $S=10^8$ and $S=10^{12}$,
with the same plots as in figure~\ref{fig:run_01}.
As with the resistive case, figure~\ref{fig:disp18} shows that the growth-rate approximately
follows the linear growth-rate until a displacement of approximately $1-2$cm (up to $t\simeq 75\tau_A$)
at which point the growth-rate of the plasma displacement slows and the profiles relax. 

A striking difference between this low-resistivity case and the resistive case presented in
section~\ref{sec:diaresist} is that the eruption no longer penetrates far into the core. Instead,
after a fast initial stage the profiles relax on timescales of $10$'s of $\tau_A$ as shown in
figure~\ref{fig:pavg18} where the surface-averages pressure profiles are plotted for the times 
indicated as vertical lines in figure~\ref{fig:disp18}.

The ELM loss $\Delta W_{ped}/W_{ped}$ is plotted as a function of time in figure~\ref{fig:run_18}
which shows a loss of $\sim 6$\% of the pedestal thermal energy (with a continuing slow relaxation of the profiles).
Most of this is due to convective
losses in the initial fast stage of the eruption as filaments are ejected from the plasma. 
Note that the simplified model used here does not include parallel heat conduction, and so only
convective losses are captured. Convective losses are thought to be important in x-point
simulations where open field-lines can quickly transport away heat, but will probably not play
such a key role in these circular ``limiter'' plasmas without open field-lines. 

\section{Effect of resistivity on ELM size}
\label{sec:elmsize}

\begin{figure}[htb!]
\centering
\includegraphics[]{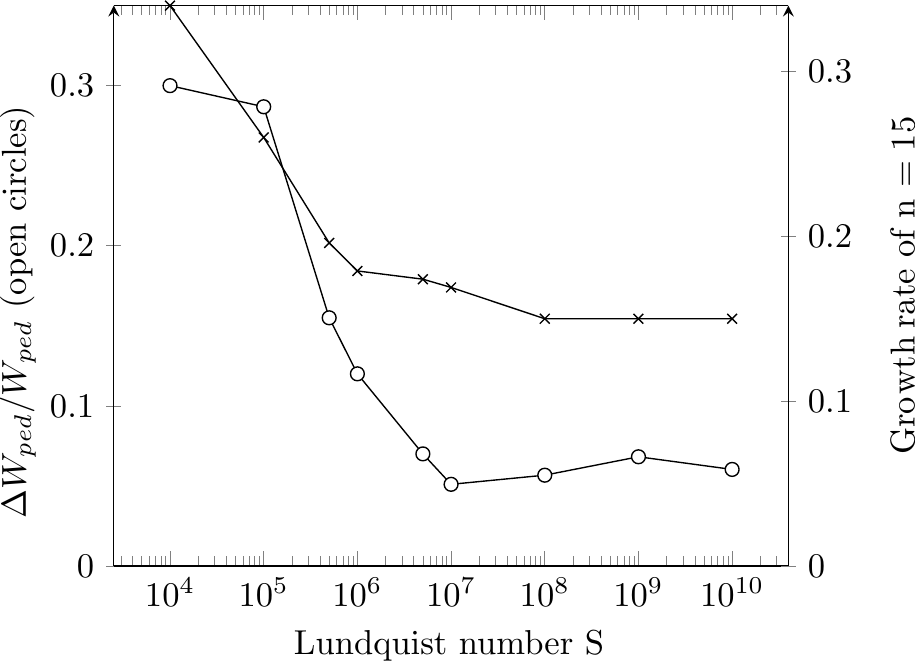}
\caption{ELM size against Lundquist number for fixed $S_H=10^{12}$}
\label{fig:elmloss}
\end{figure}

The size of the ELM eruption varies dramatically between the high and low resistivity cases
(figure~\ref{fig:run_01} and \ref{fig:run_18} respectively). The transition between these
cases is shown in figure~\ref{fig:elmloss} which plots the ELM loss fraction $\Delta W_{ped}/W_{ped}$
as a function of the Lundquist number for a fixed $S_H = 10^{12}$.
As the Lundquist number is increased, the loss of thermal energy during the ELM (open circles in
figure~\ref{fig:elmloss}) is seen to drop from
$\sim 30$\% to $\sim 6$\% over a range of $S=10^5-10^7$. One explanation for the more
violent eruptions at high resistivity is the increased linear growth-rate: Also plotted in
figure~\ref{fig:elmloss} (crosses) is the linear growth-rate of the $n=15$ mode, which also
shows an increase over a similar range of Lundquist number. 

Since there is little change in the linear growth-rate between $S=10^6$ and $S=10^7$, whilst 
ELM size changes by a factor of $\sim 2$, the process which
determines ELM loss may be a nonlinear phenomenon related to dissipation at the smallest
scales. The point at which the resistivity starts to dominate over the
hyper-resistivity ($\Delta_R=\Delta_H$) is given by $S=\sqrt{S_H\omega_A/\gamma}\sim 2\times 10^6$. Increasing the Lundquist
number above $10^6$ therefore has little effect on the ELM size $\Delta W_{ped}/W_{ped}$ because
hyper-resistivity is the dominant dissipative effect. This may then imply that in high Lundquist number
regimes relevant to high-performance discharges, the (convective) ELM loss is determined by the
hyper-resistive dissipation rather than the resistivity.

\section{Magnetic field structure}
\label{sec:magfield}

\begin{figure}[htbp!]
\centering
\subfigure[$t=60\tau_A$]{
  \label{fig:mag60}
  \includegraphics[scale=0.4]{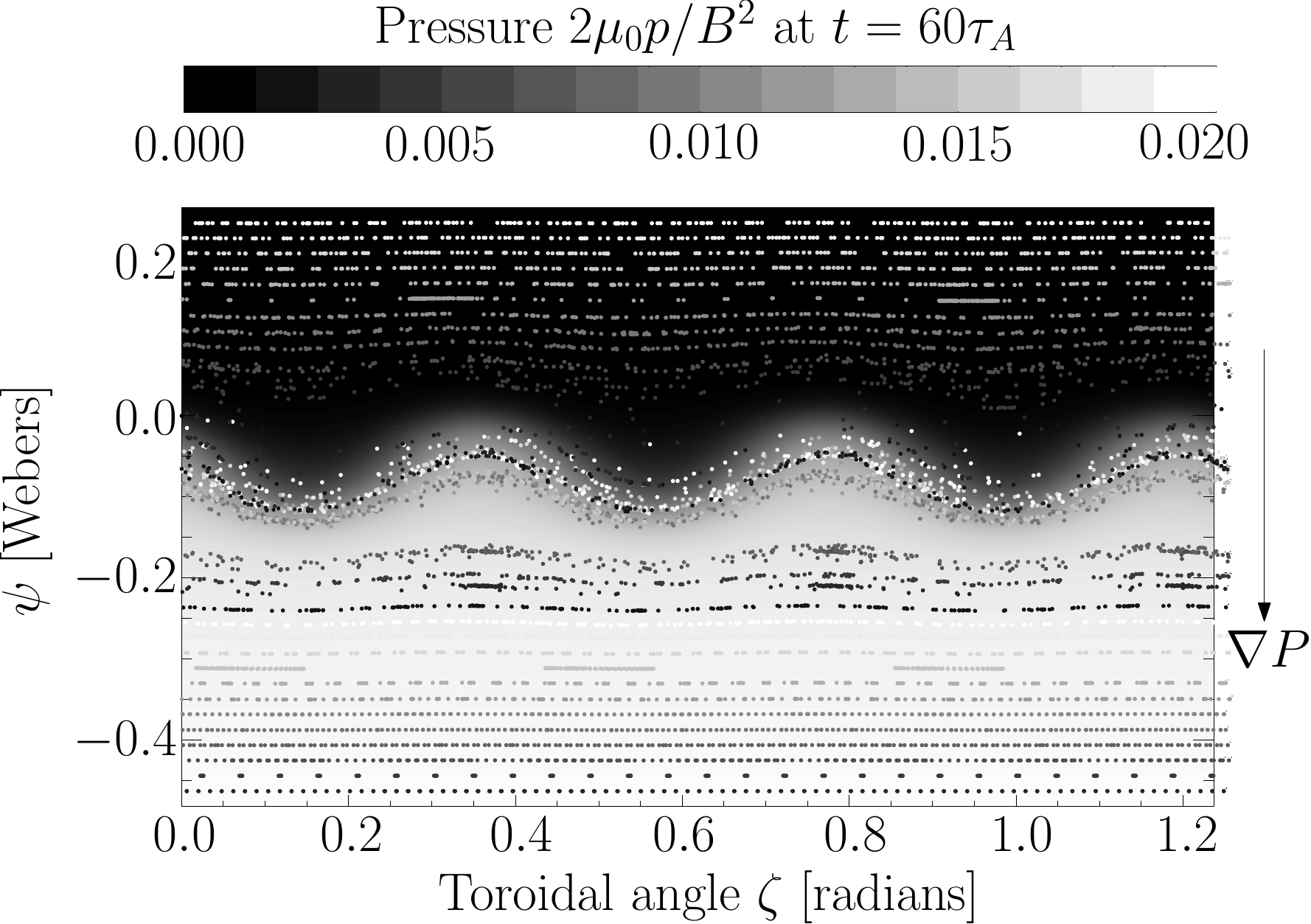}
}
\subfigure[$t=70\tau_A$]{
  \label{fig:mag70}
  \includegraphics[scale=0.4]{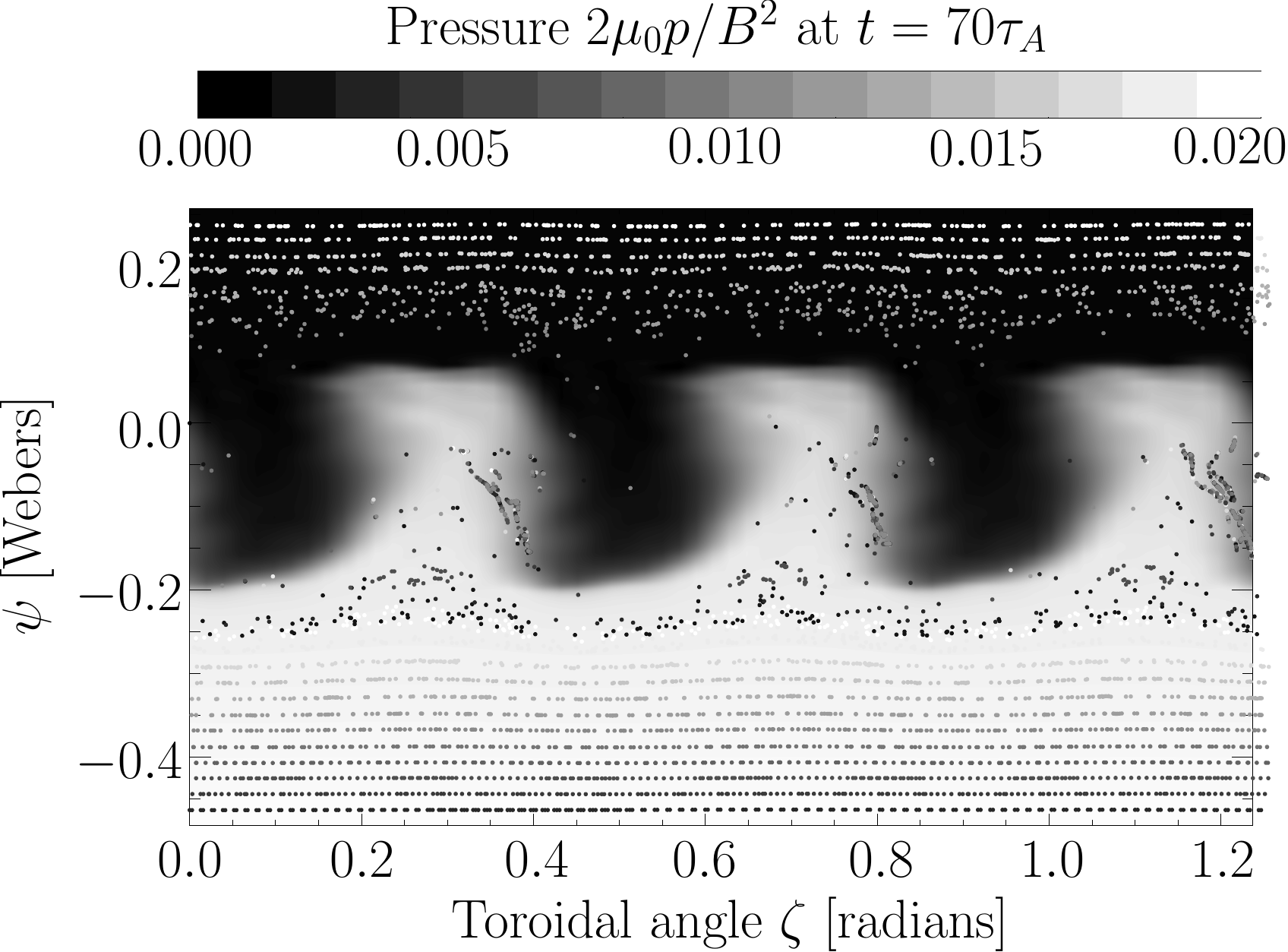}
}
\caption{Pressure and magnetic field puncture plot at outboard mid-plane. Minor radius increases going up the figures}
\label{fig:mag_18}
\end{figure}

In order to understand the ELM eruption in the low resistivity case (figure~\ref{fig:run_18}), 
Poincar\'e (puncture) plots of the magnetic field structure have been produced and compared
against the pressure contours. Figure~\ref{fig:mag_18} shows the (normalised)
pressure $2\mu_0 p/B^2$ at the outboard mid-plane as a function of $\psi$ and toroidal angle $\zeta$ for
$t=60\tau_A$ and $t=70\tau_A$. On top of these are plotted a puncture plot of the magnetic field
showing the field structure at these two times.
Up to $t\sim 60-65\tau_A$, the magnetic field surfaces follow pressure isosurfaces (frozen-in
condition), and the growth-rate remains approximately the same as the linear mode (figure~\ref{fig:disp18}).
During this period, the leading edge of the erupting filaments steepens. This forces flux-surfaces
closer together at the tips of the filaments, reducing perpendicular scale lengths and increasing the
importance of (hyper-)resistivity. Eventually, the flux-surfaces break and the magnetic
field becomes disordered. After this time ($t\simeq 65\tau_A$), the eruption growth-rate falls and
the profiles begin to relax on a slower timescale.

\section{Conclusions}

Simulations of Edge Localised Modes (ELMs) have been performed using the BOUT++
simulation code \cite{Dudson2009} for a shifted circle equilibrium (\texttt{cbm18\_dens8}).
The linear structure and growth-rates of the unstable ballooning modes have been benchmarked without
diamagnetic effects against the linear ideal MHD codes ELITE and GATO showing good agreement.
By incorporating diamagnetic drift and hyper-resistivity, the scale length of $J_{||}$ structures
can be resolved, allowing nonlinear simulations with resistivities relevant to high-performance H-mode
tokamak edge plasmas. These show reasonable ELM losses of $\sim 6$\% of the pedestal thermal energy,
with a limited radial extent, in contrast to simulations with enhanced resistivities. 

Results at high Lundquist number indicate that eruptions consist of two stages: a fast emergence
of plasma ``fingers'' at approximately the linear growth-rate (figure~\ref{fig:disp18}), during which time
the frozen-in condition is obeyed (figure~\ref{fig:mag60}). This eruption results in a steepening of the 
pressure profile at the tips of the plasma fingers, forcing flux surfaces closer together and steepening
local pressure gradients. Once
the fingers have emerged $\sim 1-2$cm, reconnection occurs at these high-gradient regions, the growth-rate
is reduced and filaments begin to break up as the plasma profiles relax (figure~\ref{fig:pavg18}).

Experimentally, it is found that the ELM size $\Delta W_{ped}/W_{ped}$ decreases as collisionality $\nu^*$
is increased~\cite{doyle2008,loarte2003}. The results presented in figure~\ref{fig:elmloss} indicate
the opposite trend at low Lundquist numbers, with the size of ELM eruptions increasing with collisionality
given the same starting pressure and $J_{||}$ profiles. At high Lundquist numbers (i.e. low $\nu^*$) 
relevant to H-mode pedestals, ELM size has been found to be quite insensitive to Lundquist number,
with dissipation dominated by hyper-resistivity. The variation of hyper-resistivity with collisionality
is however not well known. Assuming a constant electron viscosity $\mu_e\sim 1$m$^2$/s gives $\alpha_H\propto 1/\nu$, and so $S_H\sim const$ (since $S\propto 1/\nu$). This may imply that convective ELM losses
are approximately constant with collisionality, and that the observed variation of ELM loss with
$\nu^*$ is due to another process such as increasing parallel conductive losses at low collisionality.

A important consideration when interpreting the results presented here is that in these simulations,
variation in H-mode pedestal characteristics or bootstrap current with collisionality has
not been included, though in practice all these things would vary and change the point at which an ELM
is triggered. Predictions of ELM size in future devices depends on a model of the H-mode
pedestal from which to start a nonlinear ELM simulation, either coupling kinetic and fluid models
such as work coupling XGC0 to M3D~\cite{park-2007}. Other possibilities include 
using a semi-heuristic model such as EPED1~\cite{snyder2009} to provide input to a starting 
equilibrium.

Further work includes the incorporation of parallel heat conduction effects in order to study
the conductive losses in conjunction with the convective losses studied here. As mentioned in
section~\ref{sec:diaresist}, non-linear generation of poloidal flows have been found to be
important in breaking off ELM filaments~\cite{huysmans2009,pamela2010} and so incorporation of
these effects into BOUT++ simulations is a priority.

\ack

This work was supported by the UK Engineering and Physical Sciences Research Council under grant
EP/H012605/1, the US DOE by LLNL under Contract DE-AC52-07NA27344, grant DE-FG03-95ER54309 at
General Atomics, EFDA Activity WP09-MHD-04/01, and computing time through EFDA-DEISA. The views and
opinions expressed herein do not necessarily reflect those of the European Commission.
The authors wish to acknowledge P.H.Diamond for pointing out the role of hyper-resistivity in Ohm's law.

\section*{References}
\bibliography{bout++_elm}

\begin{thebibliography}{10}

\bibitem{keilhacker1984}
M~Keilhacker, F~Becker, K~Bernhardi, et~al.
\newblock {\em Plasma Phys. Control. Fusion}, 26:49, 1984.

\bibitem{doyle2008}
E~J Doyle et~al.
\newblock In {\em 22$^{nd}$ {IAEA} Fusion Energy Conference}, 2008.

\bibitem{snyder-2002}
P~B Snyder et~al.
\newblock {\em Physics of Plasmas}, 9:2037, 2002.

\bibitem{wilson-2002}
H~R Wilson et~al.
\newblock {\em Physics of Plasmas}, 9:1277, April 2002.

\bibitem{bernard1981}
L~C Bernard, F~J Helton, and R~W Moore.
\newblock {\em Comp. Phys. Comm.}, 24(3-4):377--380, 1981.

\bibitem{mikhailovskii1997}
A~B Mikhailovskii et~al.
\newblock {\em Plasma Phys. Rep.}, 23:844, 1997.

\bibitem{chapman2006}
I~T Chapman et~al.
\newblock {\em Physics of Plasmas}, 13:062511, 2006.

\bibitem{hegna-1996}
C~C Hegna, J~W Connor, R~J Hastie, and H~R Wilson.
\newblock {\em Physics of Plasmas}, 3:584, 1996.

\bibitem{connor-1998}
J~W Connor, R~J Hastie, H~R Wilson, and R~L Miller.
\newblock {\em Physics of Plasmas}, 5:2687, 1998.

\bibitem{wilson-2004}
H~R Wilson and S~C Cowley.
\newblock {\em Phys. Rev. Lett.}, 92:175006, 2004.

\bibitem{zhu-2006}
P~Zhu, C~C Hegna, and C~R Sovinec.
\newblock {\em Physics of Plasmas}, 13:102307, 2006.

\bibitem{zhu-2007}
P~Zhu et~al.
\newblock {\em Physics of Plasmas}, 14:055903, 2007.

\bibitem{sovinec-2004}
C~R Sovinec et~al.
\newblock {\em J. Comput. Phys.}, 195:355--386, 2004.

\bibitem{brennan-2006}
D~P Brennan et~al.
\newblock {\em Journal of Physics: Conference Series}, 46:63--72, 2006.

\bibitem{pankin-2007}
A~Y Pankin et~al.
\newblock {\em Plasma Phys. Control. Fusion}, 49:S63--S75, 2007.

\bibitem{burke2010}
B~J Burke et~al.
\newblock {\em Physics of Plasmas}, 17:032103, 2010.

\bibitem{snyder-2005}
P~B Snyder, H~R Wilson, and X~Q Xu.
\newblock {\em Physics of Plasmas}, 12:056115, May 2005.

\bibitem{snyder-2006}
P~B Snyder, H~R Wilson, and X~Q Xu.
\newblock In {\em {Fluid Modelling of ELMs}}, Boulder, CO, 2006.

\bibitem{huysmans-2007}
G~T~A Huysmans and O~Czarny.
\newblock {\em Nucl. Fusion}, 47:659--666, 2007.

\bibitem{huysmans2009}
G~T~A Huysmans et~al.
\newblock {\em Plasma Phys. Control. Fusion}, 51:124012, 2009.

\bibitem{pamela2010}
S~Pamela, G~Huysmans, and S~Benkadda.
\newblock {\em Plasma Phys. Control. Fusion}, 52:075006, 2010.

\bibitem{scott-2005}
B~Scott.
\newblock {\em Physics of Plasmas}, 12:102307, 2005.

\bibitem{scott-2006}
B~Scott.
\newblock {\em Plasma Phys. Control. Fusion}, 48:A387, 2006.

\bibitem{park-1999}
W~Park et~al.
\newblock {\em Physics of Plasmas}, 6:1796, 1999.

\bibitem{sugiyama-2006}
L~E Sugiyama et~al.
\newblock {\em APS Meeting}, October 2006.

\bibitem{sugiyama2009}
L~E Sugiyama and {the M3D team}.
\newblock {\em J. Phys.: Conf. Ser.}, 180:012060, 2009.

\bibitem{Ferraro2009}
N~M Ferraro et~al.
\newblock In {\em 51st {APS} meeting, November 2009}.

\bibitem{Dudson2009}
B~D Dudson et~al.
\newblock {\em Comp. Phys. Comm.}, 180:1467--1480, 2009.

\bibitem{xu2010}
X~Q Xu et~al.
\newblock {\em Submitted to Phys. Rev. Lett.}, 2010.

\bibitem{hazeltine-2003}
R~D Hazeltine and J~D Meiss.
\newblock {\em {Plasma Confinement}}.
\newblock Dover publications, 2003.

\bibitem{dimits-1993}
A~M Dimits.
\newblock {\em Phys. Rev. E}, 48(5):4070--4079, Nov 1993.

\bibitem{scott-2002}
B~D Scott.
\newblock {\em New J. Physics}, 4:52.1--52.30, July 2002.

\bibitem{Bondeson1992}
A~Bondeson, G~Vlad, and H~Lutjens.
\newblock {\em Phys. Fluids B}, 4:1889, 1992.

\bibitem{strauss1995}
H~R Strauss et~al.
\newblock {\em Physics of Plasmas}, 2:1229, 1995.

\bibitem{kaw1979}
P~K Kaw et~al.
\newblock {\em Phys. Rev. Lett.}, 43:1398, 1979.

\bibitem{caunt-2001}
S~E Caunt and M~J Korpi.
\newblock {\em A \& A}, 369:706--728, 2001.
\newblock arXiv:astro-ph/0102068.

\bibitem{loarte2003}
A~Loarte et~al.
\newblock {\em Plasma Phys. Control. Fusion}, 45:154969, 2003.

\bibitem{park-2007}
G~Park et~al.
\newblock {\em Journal of Physics: Conference Series}, 78:012087, 2007.

\bibitem{snyder2009}
P~B Snyder et~al.
\newblock {\em Nucl. Fusion}, 49:085035, 2009.

\end{thebibliography}
\bibliographystyle{unsrt}

\end{document}